\documentclass[pre,aps,showpacs,twocolumn]{revtex4-1}
\usepackage{graphicx}
\usepackage{pstricks}
\usepackage{amssymb}
\usepackage{amsmath}
\usepackage{amsfonts}
\usepackage{natbib}
\usepackage{bm}      
\usepackage{color}

\newcommand{\beq}{\begin{equation}}
\newcommand{\eeq}{\end{equation}}
\newcommand{\bea}{\begin{eqnarray}}
\newcommand{\eea}{\end{eqnarray}}
\newcommand{\ra}{\rightarrow}
\newcommand{\s}{\sigma}
\newcommand{\oh}{\omega}
\newcommand{\D}{\mathrm{d}}
\newcommand{\J}{\mathcal{J}}

\renewcommand{\r}{\right \rangle}
\renewcommand{\l}{\left \langle}

\newcommand{\m}[1]{\left \langle #1  \right \rangle}
\renewcommand{\arraystretch}{1}

\begin{document}

\title{Universal theory of efficiency fluctuations}

\author{Gatien Verley$^1$, Tim Willaert$^2$, Christian Van den Broeck$^2$, Massimiliano Esposito$^1$}
\affiliation{$^1$ Complex Systems and Statistical Mechanics, University of Luxembourg, L-1511 Luxembourg, G.D. Luxembourg}
\affiliation{$^2$ Hasselt University, B-3590 Diepenbeek, Belgium}

\date{\today}

\begin{abstract}
Using the fluctuation theorem supplemented with geometric arguments, we derive universal features of the (long-time) efficiency fluctuations for thermal and isothermal machines operating under steady or periodic driving, close or far from equilibrium. In particular, the long-time probability for observing a reversible efficiency in a given engine is identical to that for the same engine working under the time-reversed driving. When the driving is symmetric, this reversible efficiency becomes the least probable.  
\end{abstract}
\maketitle

\section{Introduction}

A thermodynamic machine is designed to operate for a given purpose such as producing mechanical work or cooling. This statement is correct ``in average'', meaning that fluctuations might occasionally prevent the machine to operate in the same way as the average behavior. While fluctuations are negligible when considering macroscopic machines, they become highly relevant at small scales when considering molecular machines or nano-devices. The macroscopic efficiency $\bar{\eta}$ used to characterize the performance of a machine ``in average'' is defined as a the ratio between an average output and an average input contribution to the macroscopic entropy production (EP) of the machine. A central result in macroscopic thermodynamics is that the second law imposes an upper bound to this macroscopic efficiency which is only reached when the machine operates reversibly. Stochastic thermodynamics has taught us that entropy production, and thus its output and input contributions, can be defined at the level of single stochastic trajectories \cite{Seifert2005_vol95, Esposito2010_vol104, Verley2012_vol108,Sekimoto2010_vol, Sinitsyn2011_vol44, Campisi2014_vol47}. In this paper we study the statistical properties of stochastic efficiencies defined at the trajectory level as ratios between such output and input. Such efficiencies may be negative or higher than the reversible efficiency, corresponding to large fluctuations along which the machine does not operate in the same mode as the average behavior. However, the fluctuation theorem \cite{Seifert2012_vol75, VandenBroeck2014_vol, Campisi2011_vol83, Esposito2009_vol81} imposes constraints on their probability distribution, more precisely on their large deviation function (LDF). Remarkably, the shape of the efficiency LDF is quites generic and displays universal features. In particular, the long-time probability for observing a reversible efficiency is identical to that of the same machine working under the time-reversed driving. The present work supersedes and completes a previous study \cite{Verley2014_vola} of machines operating at steady states or under time-symmetric drivings where it was observed that the reversible efficiency also becomes the least probable. In this paper, besides presenting results valid for general cyclic driving, we also provide an efficient method to calculate the efficiency LDF and extensively illustrate our results on a driven two-level system, see also \cite{Gingrich2014_vol} which appeared while finalizing this paper.

After framing the basic issue in section \ref{sec:EnergyConverstionNano}, the general theory and the main results are derived in section \ref{GenTheory}. Model-system illustrations of these are provided in section \ref{Model} and conclusions are drawn in section \ref{Conc}. 

\section{Thermodynamics of nanomachines} \label{sec:EnergyConverstionNano}

To study efficiency fluctuations of small-scale machines, we consider the following generic set-up. A small-scale machine is subjected to two thermodynamic forces $A_1$ and $A_2$ inducing, over a certain time $t$, the conjugated time-integrated currents  $\J_1$ and $\J_2$. While thermodynamic forces are expressed in terms of non-fluctuating properties of the macroscopic reservoirs, the currents, and hence also the efficiency, will typically fluctuate. These currents induce a fluctuating EP $\Sigma = \Sigma_1  + \Sigma_2 + \Delta S$, where $\Sigma_1 = A_1 \J_1$, $\Sigma_2 = A_2 \J_2$, and $\Delta S$ is the stochastic entropy change of the machine itself. Integrated currents and EP are time-extensive (i.e. over long times they typically grow and their average increases linearly with time). We consider small machines with finite state space meaning that the entropy changes $\Delta S$ become negligible in the long time limit (they can be shown to vanish in a large deviation sense). Their EP over long times thus reads 
\beq \label{eq:EntProds}
\Sigma \sim \J_1  A_1+ \J_2  A_2.
\eeq
Machines operate either steadily or cyclically with period $\tau$. In this latter case, time is expressed in terms of the number of period $n$ as $t = n \tau$. We denote the time-intensive variables by a lower case, e.g., $j_1=\J_1/t$ or $j_2= \J_2/t$, $\s=\Sigma/t$, $\s_1=\Sigma_1/t$ or $\s_2=\Sigma_2/t$, and ensemble averages by brackets $\l.. \r$.

A device operating as a machine (in average) uses a fueling process (the input) flowing in the spontaneous direction of its corresponding forces $\l \sigma_1 \r > 0$ (e.g. a heat flowing down a temperature gradient or particle flowing down a chemical potential gradient) in order to power a second process (the output) flowing against the spontaneous direction of its corresponding forces $\l \sigma_2 \r < 0$ (e.g. a particles flowing up a chemical potential gradient or a coordinate moving against the direction of a mechanical force). Since the second law imposes $\l \sigma \r \geq 0$, the (``type II'' \cite{Bejan2006_vol}) macroscopic efficiency of the machine defined as    
\beq
\bar \eta =-\frac{\m{\sigma_2}}{\m{\sigma_1}} = - \frac{A_2 \m{j_2} }{ A_1 \m{j_1} } \leq 1
\eeq
is always bounded upwards by the reversible (or Carnot) efficiency $\bar \eta_\mathrm{rev} =1$ occurring when $\l \sigma \r \to 0$. 
We note that traditional efficiencies (``type I'' \cite{Bejan2006_vol}) can be trivially recovered from these efficiencies. 

In the same spirit, we introduce the following time-intensive dimensionless quantity, called the stochastic efficiency:
\beq
\eta =-\frac{\sigma_2}{\sigma_1} =- \frac{ A_2 j_2 }{ A_1 j_1 } .
\eeq
As we will see below, its most probable value converges in the long time-limit to its macroscopic value $\eta \to \bar \eta$. To investigate the approach to this limit, we focus on the convergence of the intensive EPs $\sigma_1 $ and $\sigma_2$ to their most probable value, which also coincides with their average. Large deviation theory describes the exponential decay of the probability $P_t(\s_1,\s_2)$ for observing non-typical (i.e. different from their infinite-time average) EPs
\beq
\label{eq:ldcurrents} 
P_t(\s_1,\s_2) \asymp \exp\{-t I(\s_1,\s_2)\}.
\eeq
The rate $I(\s_1,\s_2)$ is the non-negative and convex LDF which reaches its minimum value at the point $\s_1=\m{\s_1}$ and $\s_2=\m{\s_2}$ which carries the entire probability weight $I(\m{\s_1},\m{\s_2}) = 0$.
 
A central result in stochastic thermodynamics known as the fluctuation theorem states that the probability to observe a positive EP in a driven machine is exponentially more likely then that of observing its negative counterpart when the machine is subjected to the time-reversed driving. This result implies that any decomposition of the EP into sub-parts that are anti-symmetric under time reversal (as is the case for $\Sigma = \Sigma_1 +\Sigma_2$) inherits the same symmetry property \cite{Garcia-Garcia2010_vol82,Garcia-Garcia2012_vol2012}. Expressed in the framework of large deviation theory, the fluctuation theorem takes the following form (we set $k_B=1$):
\beq
I(\s_1,\s_2) - \hat I(-\s_1,-\s_2) = - \s_1 - \s_2, \label{eq:jointFT}
\eeq
where $\hat P_t(\s_1,\s_2) \asymp \exp\{-t \hat I(\s_1,\s_2)\}$ is the probability of the EPs $\s_1$ and $\s_2$ for the machine working with the time reversed driving cycle. For symmetric driving cycles or for steady machines, we obviously have that $I(\s_1,\s_2) = \hat I(\s_1,\s_2)$.
In the next sections, we study the implications of this result for efficiency fluctuations.

\section{Efficiency fluctuations}\label{GenTheory}

\subsection{Large deviation function of efficiency}

We start by deriving the efficiency LDF from the EPs LDF $I(\s_1,\s_2)$.
The probability to observe an efficiency $\eta$ is given by
\bea
P_t(\eta)& = &\int \, \D \s_1 \, \D \s_2 \, P_t(\s_1,\s_2) \delta \left( \eta  + \frac{\s_2}{\s_1} \right) \nonumber \\
&= & \int \D \s_1 \,  P_t (\s_1,-\eta \s_1 ) \left |\s_1 \right | \label{eq:exactPeta}.
\eea
Inserting (\ref{eq:ldcurrents}) into (\ref{eq:exactPeta}) and using the Laplace approximation to compute the integral exactly in the long time limit, we find that 
\beq
P_t(\eta) \asymp \exp\{- t  J(\eta)\},
\eeq
where
\beq
J(\eta) = \min_{\s_1} I(\s_1,-\eta \s_1) \label{eq:defLDFeta}.
\eeq
This  result implies that  $J(\eta) \le I(0,0), \forall \eta$, an important property to be used in the next section.
Furthermore it follows that the efficiency LDF $J(\eta)$ vanishes at the macroscopic efficiency $\bar \eta$, thus corresponding to the most probable efficiency:
\beq
J\left( \bar \eta \right) = \min_{\s_1} I\left(\s_1, \s_1\frac{\langle \s_2 \rangle}{\langle \s_1 \rangle} \right)=0.
\eeq
The minimum value zero is reached for $\s_1 = \langle \s_1 \rangle$ because the LDF for EPs is convex and vanishes at the average EPs.

\subsection{Geometric interpretation}
\label{sec:GeometricInterpretation}

\begin{figure}
\vspace{0.4cm}
\begin{center}
\includegraphics[width=7cm]{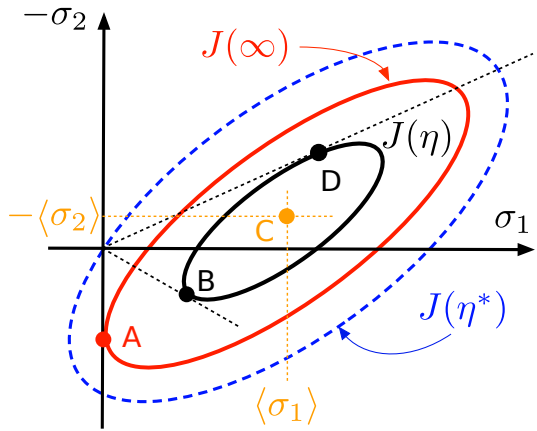}
\end{center}
\caption{Typical contour lines of the LDF $I(\s_1,\s_2)$. The point C corresponds to the most probable value $I(\m{\s_1},\m{\s_2})=0$. A straight line through the origin with slope $\eta_D$ touches the contour line, whose $I$-value equals $J(\eta_D)$ (idem for point B sharing the same $J$ value). The maximum of $J(\eta)$ corresponds to $I$-value of the contour crossing the origin $J(\eta^*)=I(0,0)$ (blue long dashed line), while $J( \infty)$ to that of the contour touching the $\s_2$-axis in A (red solid line). \label{fig1}}
\end{figure}

\begin{figure}
\vspace{0.4cm}
\begin{center}
\includegraphics[width=7cm]{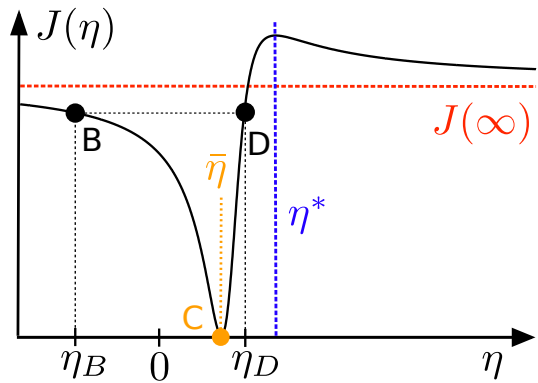}
\end{center}
\caption{Typical shape of the efficiency LDF $J(\eta)$. For steady state machines or machines with time-symmetric driving cycles, the shape is the same and the maximum is at the reversible efficiency $\eta^* = \bar \eta_\mathrm{rev}=1$. The horizontal asymptote corresponds to the point A of Fig.~\ref{fig1}.  \label{fig5}}
\end{figure}

We now analyze the contour lines of the LDF $I(\s_1,\s_2)$ in the input and output $(\s_1,-\s_2)$ plane represented in Fig.~\ref{fig1}. They form closed convex lines encircling the point C=$(\m{\s_1},\m{\s_2})$, where $I$ reaches its minimal value $I=0$, see Fig.~\ref{fig1}. This point corresponds to the most probable efficiency $\eta=\bar \eta$ and must lie in the upper right corner of the $(\s_1,-\s_2)$ plane for the device to operate as a machine. 
A given value of the stochastic efficiency corresponds to a straight line with slope $\eta$ and crossing the origin: $-\s_2=\eta \s_1$. The corresponding value of the LDF $J(\eta)$ is, according to (\ref{eq:defLDFeta}), found as the minimum of $I(\s_1,\s_2)$ along this line. This minimum is reached for the contour line closest to the most probable point C, namely the contour tangent to the line of slope $\eta$. 
A given contour line has two tangent lines crossing the origin and corresponding to two different efficiencies with the same value of the LDF. For instance, in Fig.~\ref{fig1} the black solid contour line has two tangent lines, one in B and one in D, corresponding to the efficiencies $\eta_B$ and $\eta_D$ and to the same value of the LDF $J(\eta_B) = J(\eta_D)$ as shown in Fig~\ref{fig5}. 

From the above geometric analysis of Eq.~(\ref{eq:defLDFeta}) illustrated in Fig~\ref{fig1}, we can deduce the overall typical shape of the efficiency LDF $J(\eta)$ represented in Fig.~\ref{fig5}. Starting from the point C with efficiency $\eta=\bar \eta$ and decreasing the slope $\eta$, $J(\eta)$ increases until the contour line touches the vertical axis in point A, with the corresponding efficiency $\eta=-\infty$. Similarly, increasing the slope $\eta$ from C upward, the LDF $J(\eta)$ increases until the contour line crosses the origin corresponding to $I(0,0) = J(\eta^*)$ where $\eta^*$ is the contour slope at the origin. This efficiency corresponds, as shown above, to the maximum value of $J(\eta)$, hence $\eta^*$ is the least probable efficiency in the sense of large deviations. For $\eta\ge \eta^*$, the intersection between the contour and the efficiency line moves from the upper right corner to the lower left corner of the plane, and the LDF decreases until its limiting value is reached for $\eta=+\infty$. Positive and negative infinite efficiencies share the same contour line touching the vertical axis in A, with the same limiting $J(\infty)$-value. 


\subsection{Least likely and reversible efficiency}
\label{sec:AsymDriving}

We have shown that the least probable efficiency is given by the slope in zero of the contour line crossing the origin. Along this contour line the total differential of $I$ has to vanish
\beq
\D I = \frac{\partial I}{\partial \s_1} \D \s_1 + \frac{\partial I}{\partial \s_2} \D \s_2 = 0.
\eeq 
Evaluating this equation at the origin one gets
\beq
\eta^* = - \frac{\D \s_2}{\D \s_1} =  \frac{\partial I}{\partial \s_1}  \left( \frac{\partial I}{\partial \s_2} \right)^{-1}
\eeq
and similarly for the machine subjected to the time-reversed driving cycle 
\beq
\hat \eta^* = -\frac{\D \s_2}{\D \s_1} =  \frac{\partial \hat I}{\partial \s_1}  \left( \frac{\partial \hat I}{\partial \s_2} \right)^{-1},
\eeq
where $\eta^*$ and $\hat \eta^*$ are defined by $J(\eta^*) = I(0,0)$ and $\hat J(\hat \eta^*) = \hat I(0,0)$. Taking the partial derivative with respect to $\s_1$ and $\s_2$ of the fluctuation theorem (\ref{eq:jointFT}) and evaluating it at vanishing EPs leads to the following two equations
\beq
\frac{\partial I}{\partial \s_1} + \frac{\partial \hat I}{\partial \s_1} = -1, \qquad 
\frac{\partial I}{\partial \s_2} + \frac{\partial \hat I}{\partial \s_2} = -1. \label{eq:ParDerFT}
\eeq
Therefore, the least probable efficiency of the machine subjected to the time-reversed driving cycle is related to the EPs LDF of the original machine by
\beq
\hat \eta^* = \left( 1 + \frac{\partial I}{\partial \s_1} \right)\left( 1 + \frac{\partial I}{\partial \s_2} \right)^{-1}.
\eeq
For machines operating at steady state or subjected to time-symmetric driving cycles, $I(\s_1,\s_2)=\hat I(\s_1,\s_2)$ and from Eq.~(\ref{eq:ParDerFT}), one recovers the result first derived in Ref.\cite{Verley2014_vola} stating that the least probable efficiency is the reversible efficiency: $\eta^*=\hat \eta^* = \bar \eta_\mathrm{rev}=1$. 
However, if the machine works with non-time-symmetric cyclic driving, the reversible efficiency is not the least probable any more but remains a special point of the LDF. Indeed, if we evaluate Eq.~(\ref{eq:jointFT}) in $\s_2 = - \bar \eta_\mathrm{rev} \s_1 = - \s_1$, we find that
\beq
I(\s_1, -\bar \eta_\mathrm{rev} \s_1) = \hat I(-\s_1,\bar \eta_\mathrm{rev} \s_1), \label{eq:sym1}
\eeq
which after minimization over $\s_1$ implies that 
\beq
J(\bar \eta_\mathrm{rev}) = \hat J(\bar \eta_\mathrm{rev}). \label{Central}
\eeq
Hence, the efficiency LDF takes the same value for both machines, forward and reverse, at the value of the reversible efficiency. The two LDFs will thus cross at this point. 

\subsection{Efficiency LDF from EPs cumulant generating function}

We now propose a convenient way to calculate the efficiency LDF directly using the EPs cumulant generating function (CGF) 
\beq
\phi(\gamma_1,\gamma_2) = \lim_{t\ra \infty} \frac{1}{t} \ln \l e^{\gamma_1 \Sigma_1+\gamma_2 \Sigma_2} \r.
\eeq
This latter is typically obtained analytically or numerically from the dominant eigenvalue of a dressed stochastic generator \cite{Lebowitz1999_vol95,Gaspard2004_vol120,Esposito2007_vol76}. The CGF and LDF for EPs are known to be related by a Legendre transform \cite{Touchette2009_vol478}:
\bea
\phi(\gamma_1,\gamma_2) &=& \max_{\s_1,\s_2} \left\{ \gamma_1 \s_1 + \gamma_2 \s_2 - I(\s_1,\s_2) \right \}, \\
I(\s_1,\s_2) &=& \max_{\gamma_1,\gamma_2} \left\{ \gamma_1 \s_1 + \gamma_2 \s_2 - \phi(\gamma_1,\gamma_2) \right \}.
\eea
Therefore, the minimization of Eq.~(\ref{eq:defLDFeta}) can be directly performed on
\bea
I(\s_1,-\eta \s_1) &=& \max_{\gamma_1,\gamma_2} \left\{ (\gamma_1 - \gamma_2 \eta ) \s_1 - \phi(\gamma_1,\gamma_2) \right \}.
\eea
Using the change of variable $\gamma = \gamma_1 -\gamma_2 \eta $ and the efficiency LDF definition, we get
\beq
J(\eta) = \min_{\s_1} \left \{ \max_{\gamma } \left[   \gamma \s_1 + \max_{\gamma_2} \left( - \phi(\gamma+\gamma_2 \eta,\gamma_2) \right) \right] \right \}.
\nonumber
\eeq
Defining the function
\beq
f_\eta(\gamma) = -\max_{\gamma_2} (-\phi(\gamma+\gamma_2 \eta,\gamma_2))  = \min_{\gamma_2} \phi(\gamma+\gamma_2 \eta,\gamma_2),
\label{eq:fdef}
\eeq
whose Legendre transform is such that:
\bea
\mathcal{F}_\eta (\s_1) &=& \max_{\gamma} \left\{ \gamma \s_1 - f_\eta(\gamma) \right \}, \\
f_\eta (\gamma) &=& \max_{\s_1} \left\{ \gamma \s_1 - \mathcal{F}_\eta(\gamma) \right \},
\eea
the efficiency LDF can be written as:
\bea
J(\eta) &=& \min_{\s_1} \left \{ \max_{\gamma} \left [ \gamma \s_1  - f_\eta(\gamma) \right ]\right \} , \\
&=& \min_{\s_1} \mathcal{F}_\eta(\s_1) \nonumber, \\
&=& - \max_{\s_1} \left\{ - \mathcal{F}_\eta(\s_1)  \right\} \nonumber, \\
&=& - f_\eta(0). \nonumber
\eea
Using Eq.~(\ref{eq:fdef}) we finally conclude that
\beq
J(\eta) = - \min_{\gamma_2} \phi(\gamma_2 \eta, \gamma_2). \label{eq:Jfromphi}
\eeq
This result is of significant practical importance because it shows that the efficiency LDF can be obtained using a simple minimization procedure from the EPs CGF which can be calculated using well known conventional techniques.

\subsection{Efficiency fluctuations close to equilibrium} \label{sec:CloseEq}

close to equilibrium, the CGF of EPs is generically a quadratic function
\beq
\phi(\gamma_1,\gamma_2) = \frac{1}{2} \sum_{i,k=1,2} C_{ik} \gamma_i \gamma_k + \sum_{k=1,2} \gamma_k \l \s_k \r \label{eq:QuadPhi}
\eeq
with $C_{ik}$ the asymptotic value of the covariance matrix elements $C_{ik}(t)= (\l \Sigma_i(t) \Sigma_k (t)\r - \l  \Sigma_i(t) \r  \l \Sigma_k(t) \r)/ t $. The position of the minimum $\gamma_2^*$ in Eq.~(\ref{eq:Jfromphi}) is solution of $\D \phi(\gamma_2 \eta,\gamma_2) / \D \gamma_2 = 0$ and reads:
\beq
\gamma_2^* =  -\frac{\eta \l \s_1 \r + \l \s_2 \r}{\eta^2C_{11}+2\eta C_{12}+C_{22}}. \label{eq:PosMin}
\eeq
It follows from equation (\ref{eq:Jfromphi}), (\ref{eq:QuadPhi}) and (\ref{eq:PosMin}) that the efficiency LDF close to equilibrium is $J(\eta) = - \phi (\gamma_2^*\eta,\gamma_2^*)$, namely:
\beq
J(\eta) = \frac{1}{2} \frac{(\eta \l \s_1 \r + \l \s_2 \r)^2}{\eta^2C_{11} + 2\eta C_{12}+C_{22}}. \label{eq:closeEqJPartial}
\eeq
From linear response theory, currents are a linear combination of the affinities, $\l j_i \r = \sum_{k=1,2} L_{ik} A_k$, with the Onsager coefficient defined by
\beq
L_{ik}= \lim_{t \rightarrow \infty }\frac{1}{2t} \l [\J_i(t)-\langle \J_i \rangle_{eq} ] [ \J_{k}(t) - \langle \J_k \rangle_{eq}] \r_{eq}.
\eeq
They are related to the covariance matrix of EPs by $ \lim_{t \rightarrow \infty} C_{ik}(t)/2 = A_i L_{ik} A_k$.
Then, the average EPs are related to the asymptotic covariance matrix by $\l \s_i \r = \sum_{k=1,2} C_{ik}/2$
so that Eq.~(\ref{eq:closeEqJPartial}) can be rewritten as:
\beq
J(\eta) = \frac{1}{8} \frac{[\eta C_{11} + (1+\eta)C_{12} + C_{22}]^2}{\eta^2C_{11}  + 2\eta C_{12}   +C_{22} }.  \label{eq:LDFetaCovar}
\eeq
This relation also results from combining the fluctuation theorem (\ref{eq:jointFT}) with the Gaussian LDF of the currents,
\beq
I(\s_1\s,_2) = \frac{1}{2} \sum_{i,k=1,2} \left( C^{-1} \right)_{ik} (\s_i-\l \s_i \r)(\s_k-\l \s_k \r),
\eeq
obtained by Legendre-transforming (\ref{eq:QuadPhi}).
We note that while (\ref{eq:jointFT}) involves the EPs LDF of the machine subjected to the direct as well as to the time-reversed driving cycles. However, close to equilibrium the EPs LDF for both these drivings can be shown to follow the same statistics (see ch.3, sec. 2.3 of \cite{Verley2012_vol}).

\section{Two-state cyclic machine}\label{Model}

\label{sec:2statemodel}

\renewcommand{\arraystretch}{2.6}
\setlength{\tabcolsep}{0.4cm}
\begin{table*}
\begin{tabular}{r|cccc} 
		& Heat engine 	 	& Heat pump	& Refrigerator  \\
\hline
$\sigma_1 = j_1 \times A_1 $ & $q_h \times \displaystyle \frac{\eta_C}{T_c} $ & $w \times \displaystyle \frac{1}{T_c} $ & $w \times \displaystyle \frac{1}{T_h}$ \\
$\sigma_2 = j_2 \times A_2$   & $(-w) \times\displaystyle \left(- \frac{1}{T_c}\right) $         & $ (-q_h) \times \left(\displaystyle - \frac{\eta_C}{T_c} \right)$       & $q_c \times \left( \displaystyle - \frac{\eta_C}{T_c} \right)$   \\
$\phi(\gamma_1,\gamma_2)  $ & $ \varphi_\mathrm{h}(\gamma_2 / T_\mathrm{c}, \gamma_1 \eta_\mathrm{C} / T_\mathrm{c}) /\tau$  & $\varphi_\mathrm{h}(\gamma_1 / T_\mathrm{c},\gamma_2\eta_\mathrm{C} / T_\mathrm{c})/\tau$  & $\varphi_\mathrm{c}(\gamma_1 / T_\mathrm{h}, - \gamma_2 \eta_\mathrm{C} / T_\mathrm{c} )/\tau$ 
\end{tabular}
\caption{EPs and their connection to current and affinities for a thermal machine operating respectively as heat engine, heat pump and refrigerator. The input and output currents $j_1$ and $j_2$ are always positive in average. We use $\eta_\mathrm{C} = 1- T_\mathrm{c}/T_\mathrm{h}$. The last line indicates how to obtain $\phi$ the CGF for the EPs per unit time from the CGF of the work and heat per period.
\label{tab:ThermalMachines}}
\end{table*}
\renewcommand{\arraystretch}{1.4}
\setlength{\tabcolsep}{0.4cm}

\begin{figure}
\begin{center}
\includegraphics[width=\columnwidth]{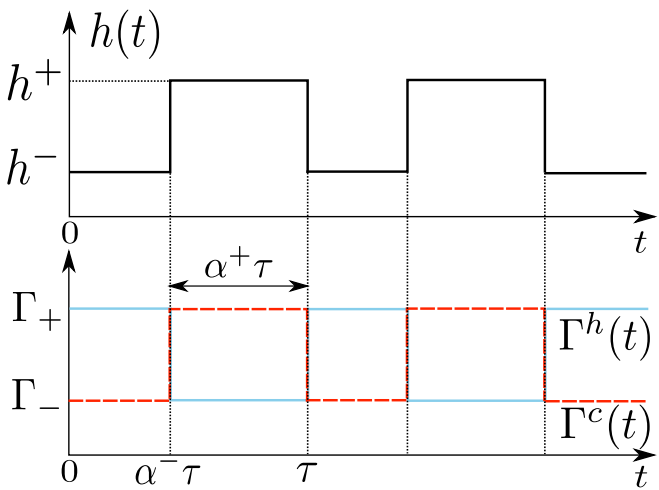}
\end{center}
\caption{(Top) External driving $h(t)$ following a piecewise constant protocol of period $\tau$. The driving takes the value $h^\pm$ during $\alpha^\pm \tau$, with $\alpha^-+\alpha^+= 1$. (Bottom) Time evolution of $\Gamma^\nu(t)$ indicating the coupling with cold reservoir $\nu=c $ for the blue solid line and with the hot reservoir $\nu = h$ for the red dashed line. At driving $h^\pm$, the hot reservoir coupling is $\Gamma_\pm$ and the cold reservoir coupling is $\Gamma_\mp$.
}
\label{fig0}
\end{figure}

To illustrate the results of the previous section, we consider a system made of two states $\s = \pm 1$ coupled to a cold and a hot heat reservoirs $\nu=c,h$ at temperatures $T_\nu = 1/\beta_\nu$. The system energies $E(t)=-h(t) \s(t)$ are modulated by an external piecewise constant driving $h(t)$ of period $\tau$, where $\s(t)$ denotes the system state at time $t$. The energy changes in the system due to changes in the driving $h(t)$ (occurring at fixed $\s$) constitute work. The energy changes between system states (occurring at a fixed driving value $h$) induced by either reservoirs and and occurring at random times constitute heat (work and heat are by convention positive when they increase the system energy). The Markovian rates describing these latter transitions from $\s$ to $-\s$ due to reservoir $\nu$ are of the form 
\beq
\quad k_{\nu}(h(t),\s) = \oh_{\nu}(h(t)) e^{-\beta_{\nu} h(t) \s},
\eeq
and thus satisfy local detailed balance. We consider Fermi rates $\oh_\nu(h(t)) = \Gamma_\nu(h(t)) / [2 \cosh \beta_{\nu}(h(t))]$, but Arrhenius rates $\oh_\nu(h(t)) = \Gamma_\nu(h(t))$ or Bose rates $\oh_\nu(h(t)) = \Gamma_\nu(h(t)) / |2\sinh \beta_{\nu}(h(t))|$ may be considered as well. 
In order for the system to operate as a thermal machine, the coupling constants $\Gamma_\nu(h(t))$ between the system and the reservoir $\nu$ have to depend on the driving value. 
The heat per period received from the hot (resp. cold) reservoir is denoted $q_h$ (resp. $q_c$). The work per period performed by the driving on the system is denoted $w$. Table \ref{tab:ThermalMachines} describes the different possible operating regimes of our thermal machine and explains how to relate their EPs and their efficiency to the general formulation of section \ref{sec:EnergyConverstionNano}.\\

\begin{figure}
\begin{center}
\includegraphics[width=\columnwidth]{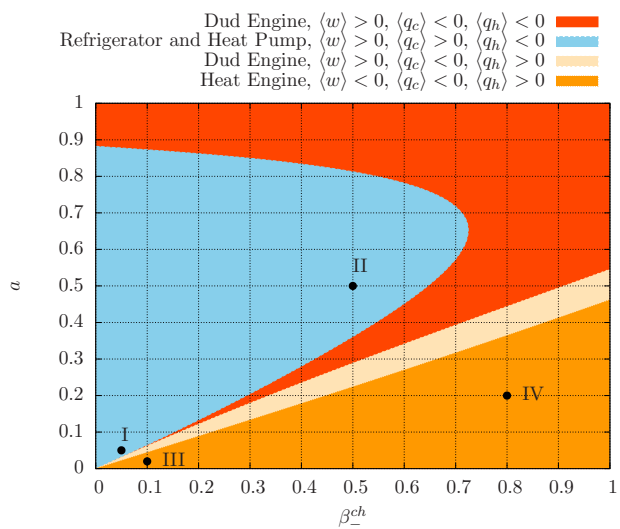}
\end{center}
\caption{Diagram representing the various operating modes of our thermal machine subjected to the driving cycle depicted in Fig. \ref{fig0}, as a function of the inverse temperature difference $\beta_-^{ch} = \beta_{c}-\beta_{h}$ and the amplitude $a$ of the driving. We used: inverse temperature $\beta_+^{ch}=\beta_c + \beta_h = 2$, period $\tau= 1$, cyclic ratio $\alpha^- = 0.3$, bare field $h_0 = 1$, coupling constants $\Gamma_- = 0.25$ and $\Gamma_+ =4$. These parameters are also used for all figures using this model (excepted Fig.~\ref{fig11}). Black circles correspond to the values of $a$ and $\beta_-^{ch}$ used in Fig.~\ref{fig12}, \ref{fig10} and \ref{fig9}. I: $\beta_-^{ch}=0.05$, $a=0.05$; II:  $\beta_-^{ch}=0.5$, $a=0.5$; III: $\beta_-^{ch}=0.1$, $a=0.02$; IV: $\beta_-^{ch}=0.8$, $a=0.2$. 
\label{fig7}}
\end{figure}

\begin{figure*}
\begin{tabular}{ccc}
\includegraphics[width=5.6cm]{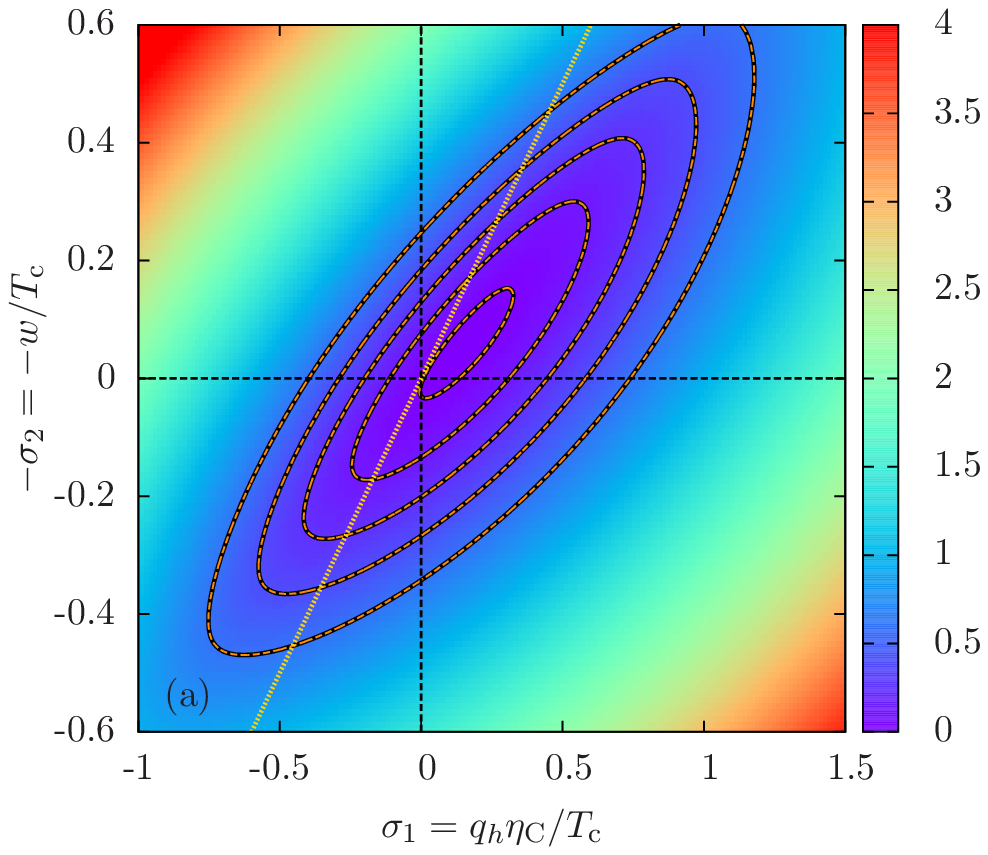} & 
\hspace{-0.6cm}\includegraphics[width=5.6cm]{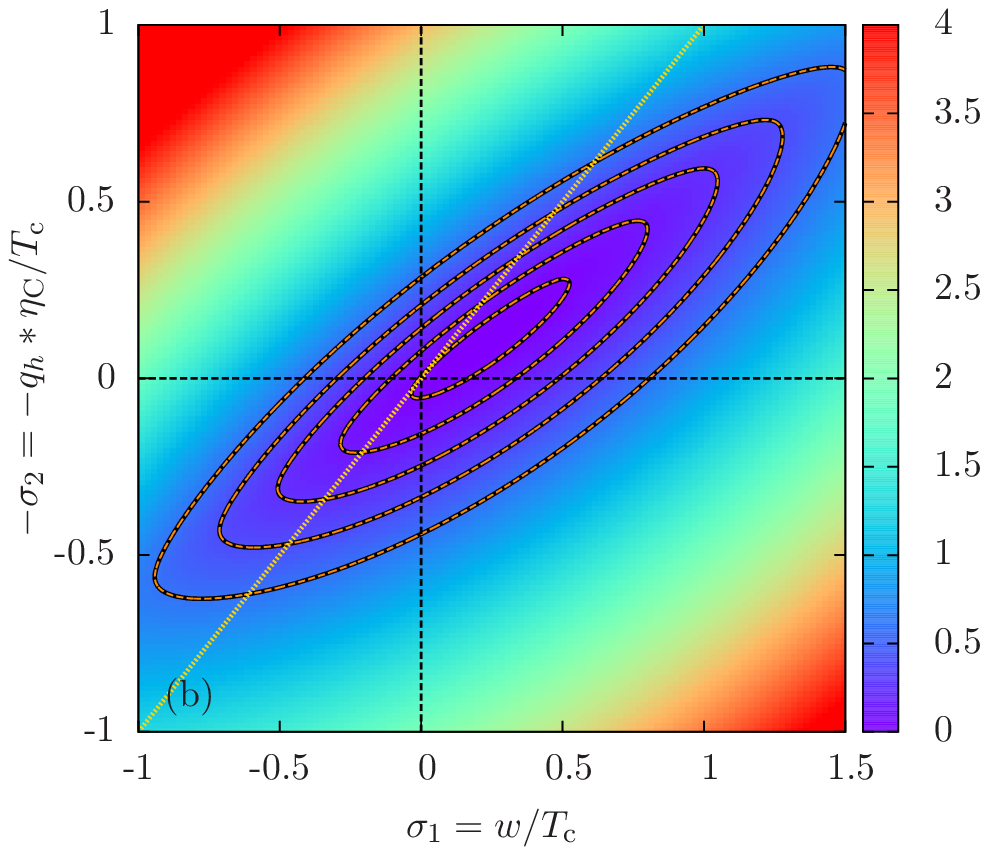} & 
\hspace{-0.6cm}\includegraphics[width=5.6cm]{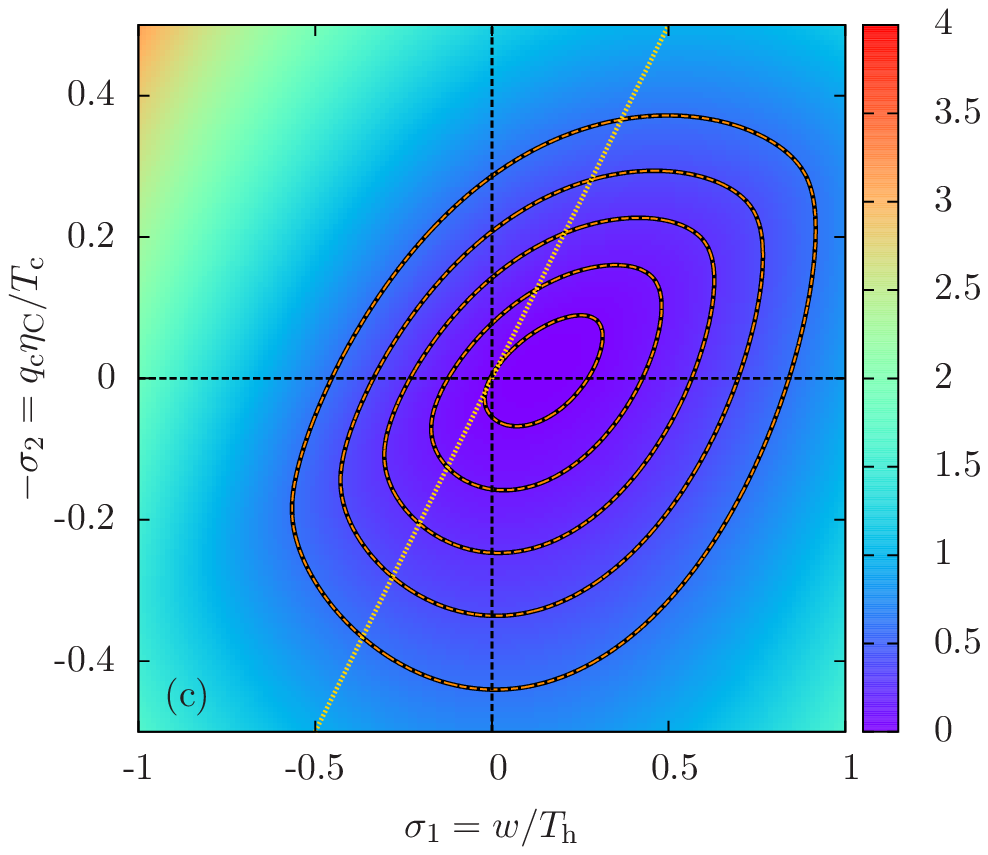} \\
\includegraphics[width=5.6cm]{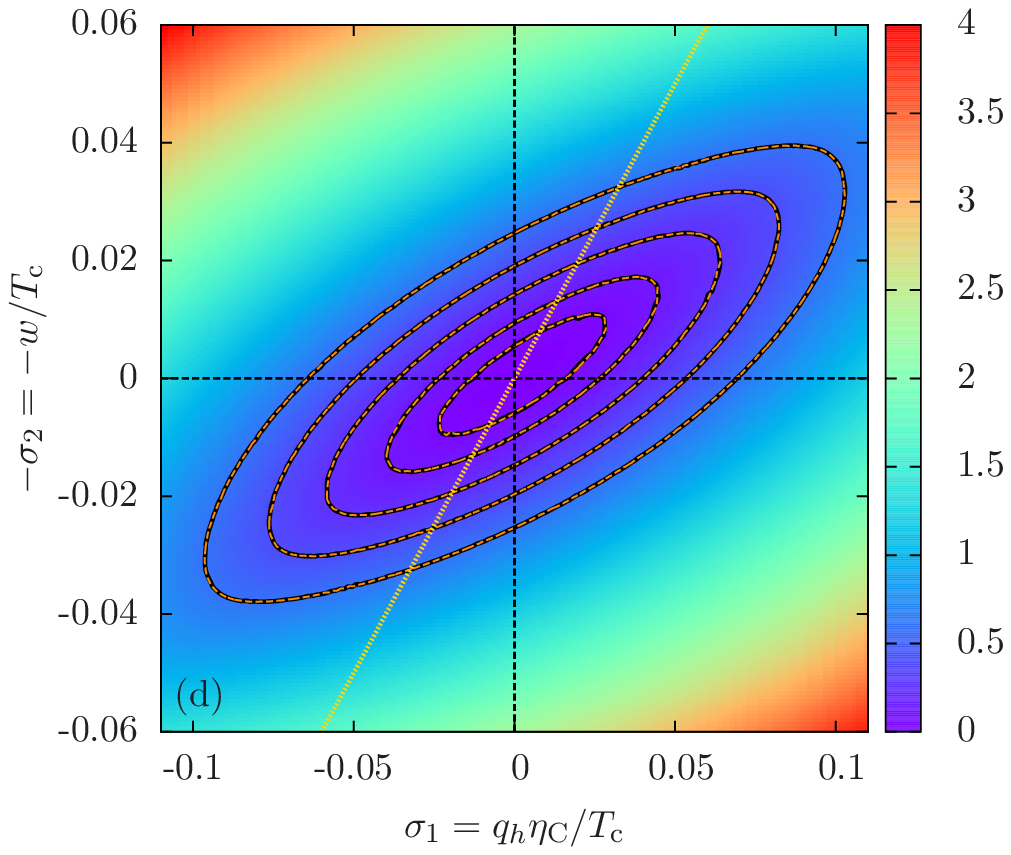} &
\hspace{-0.6cm}\includegraphics[width=5.6cm]{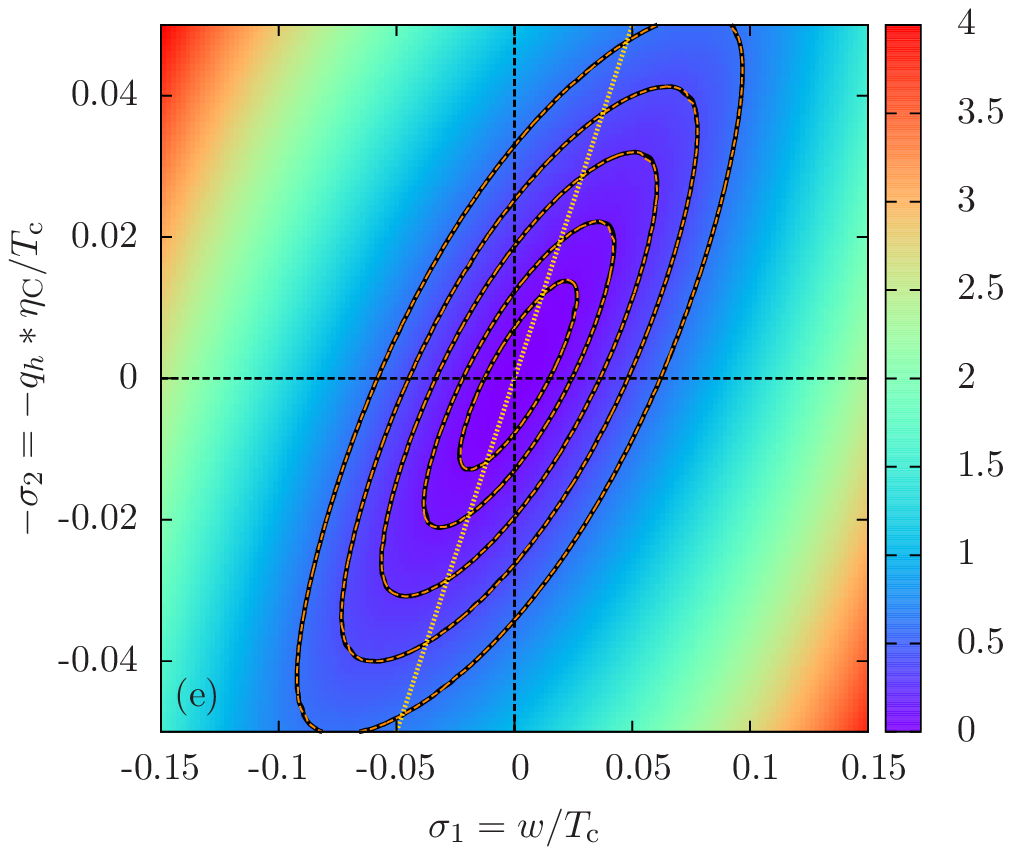} &
\hspace{-0.6cm}\includegraphics[width=5.6cm]{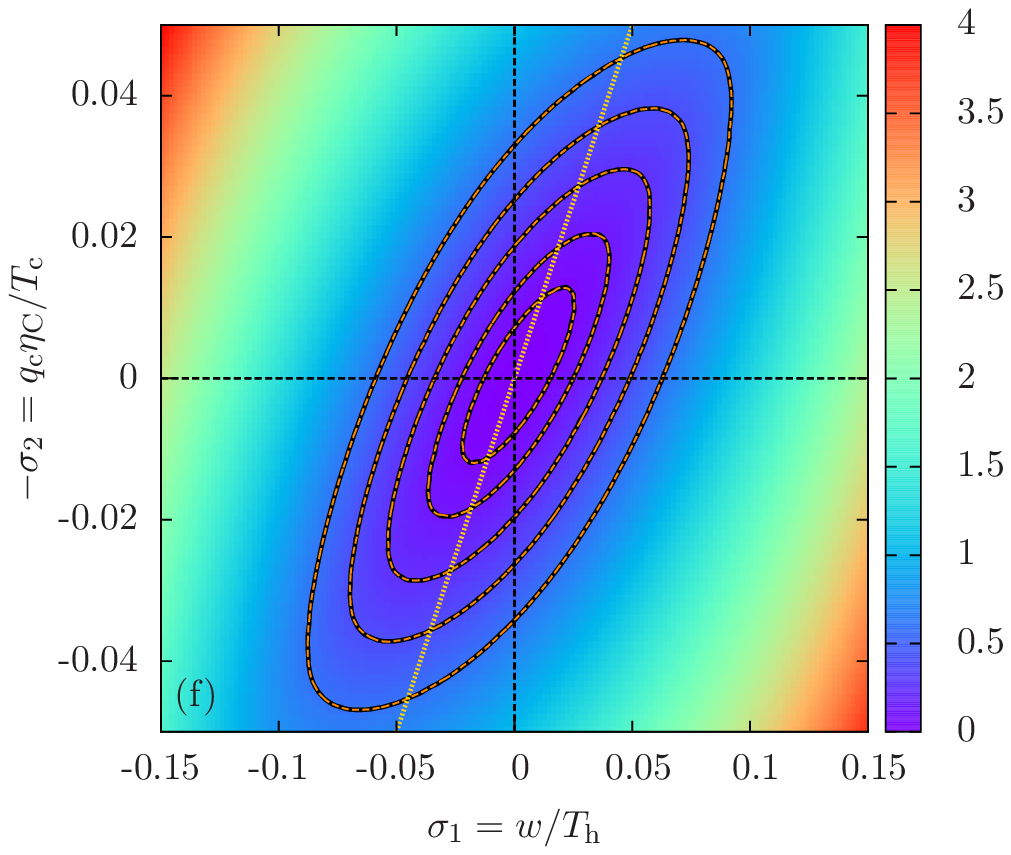}
\end{tabular}
\caption{LDF of work and heat for the three operating modes of the machine: heat engine (a,d), heat pump (b,e) and refrigerator (c,f), far-from- (top row) or close to equilibrium (bottom row). The parameter values correspond to the black points in Fig.~\ref{fig7}. The solid black level lines corresponding to $I(\s_1,\s_2)$ and the orange dashed lines corresponding $I(-\s_1,-\s_2)-\s_1-\s_2$ perfectly coincide, illustrating the fluctuation theorem symmetry for symmetric driving under time-reversal. The yellow dotted line is the straight line of slope $\bar \eta_\mathrm{rev}$ crossing the origin.\label{fig12}}
\end{figure*}

We first consider the piecewise constant driving depicted in Fig.~\ref{fig0}. 
This driving is symmetric under time-reversal (up to a time shift negligible in the long time limit) and the single reservoir version of this model was studied analytically in \cite{Verley2013_vol88, Verley2014_vol16}.
The work and heat CGF can also be calculated analytically for our machine as described in Appendix \ref{sec:CGFWQ}.

Depending on the choice of the various parameters, this machine operates in the different modes illustrated in Fig.~\ref{fig7}.
Note that refrigerators and heat pumps only differ by the way efficiency is defined: using either the heat from the cold reservoir or from the hot reservoir for output process. We also see that the heat pump/refrigerator region is separated from the heat engine region by two different dud engine regions. The red region corresponds to a heater using work to heat the hot and the cold reservoirs. The light beige region is also a dud engine that uses work to enhance the heat flow in its spontaneous direction.
The two black points in the blue as well as in the orange region correspond to the close and far-from-equilibrium parameter values. 

Using Table \ref{tab:ThermalMachines} and the exact CGF derived for work and heat in Appendix \ref{sec:CGFWQ}, the efficiency LDF $J(\eta)$ can be directly obtained by a numerical minimization as suggested by (\ref{eq:Jfromphi}). Alternatively, one could also directly minimize the LDF $I(\s_1,\s_2)$ computed via a two dimensional Legendre transform of the joint CGF. 

In Fig.~\ref{fig12} the LDF $I(\s_1,\s_2)$ is displayed for the three operating modes of the machine, both close and far from equilibrium. Since the driving depicted in Fig.~\ref{fig0} is symmetric under time-reversal, we verify that $I(\s_1,\s_2) = \hat I(\s_1,\s_2)$ as predicted by (\ref{eq:jointFT}) and the least probable efficiency coincides with the reversible efficiency (given by the slope of the level line crossing the origin). The fact that the level lines become elliptical close to equilibrium illustrates that the statistics of the EPs, $\s_1$ and $\s_2$, becomes Gaussian. 

\begin{figure}
\begin{center}
\includegraphics[width=\columnwidth]{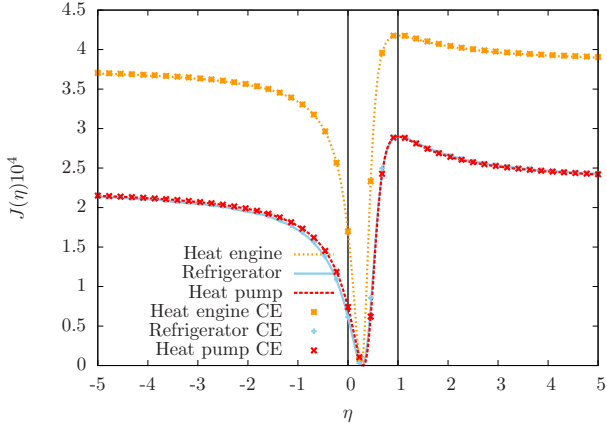}
\end{center}
\caption{efficiency LDF for the three main types of machines working close to equilibrium (CE). Lines are for $J(\eta)$ obtain from equation (\ref{eq:Jfromphi}) and table \ref{tab:ThermalMachines} and symbols come from the close-to-equilibrium prediction of equation (\ref{eq:LDFetaCovar}). The chosen parameter corresponds into Fig.~\ref{fig7} to point I for the heat pump and the refrigerator, and to point III for the heat engine. The indicated type of machine corresponds to the average behavior. \label{fig10}}
\end{figure}

\begin{figure}
\begin{center}
\includegraphics[width=\columnwidth]{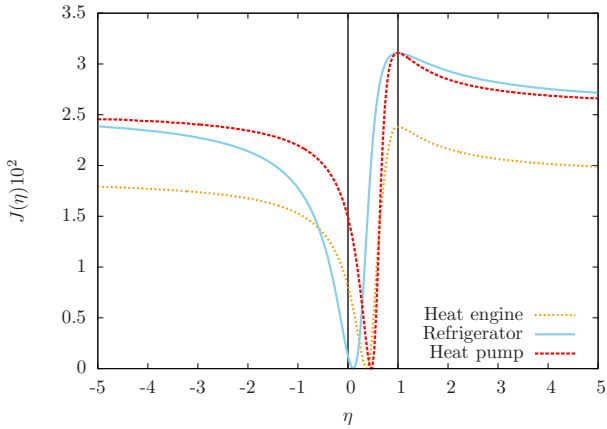}
\end{center}
\caption{efficiency LDF for the three main types of machines working far from equilibrium. The chosen parameter corresponds into Fig.~\ref{fig7} to point II for the heat pump and the refrigerator, and to point IV for the heat engine. The indicated type of machine corresponds to the average behaviour. \label{fig9}}
\end{figure}

In Fig.~\ref{fig10} (reps. Fig.~\ref{fig9}), we plot $J(\eta)$ for the three operating mode of the machine operating close to equilibrium (resp. far from equilibrium) and corresponding to the black points in Fig.~\ref{fig7}. 
We verify that the reversible efficiency corresponds to a maximum of the LDF as it should for time-symmetric drivings. Fig.~\ref{fig10}, also confirms the validity of our close-to-equilibrium theory presented in section \ref{sec:CloseEq}. 
We observe that the curves for the refrigerator and the heat pump are very similar in the close-to-equilibrium limit. 
The plateau value for large efficiencies and the value of the efficiency LDF at the reversible efficiency are also very similar on Fig.~\ref{fig9} in far-from-equilibrium conditions, even though the position of the most probable efficiency is different. 
Finally we remark that for all the parameter values and operating mode that we considered, the global shape of the efficiency LDF is consistent with the shape illustrated in Fig. \ref{fig5}.\\

\begin{figure}
\begin{center}
\includegraphics[width=8.0cm]{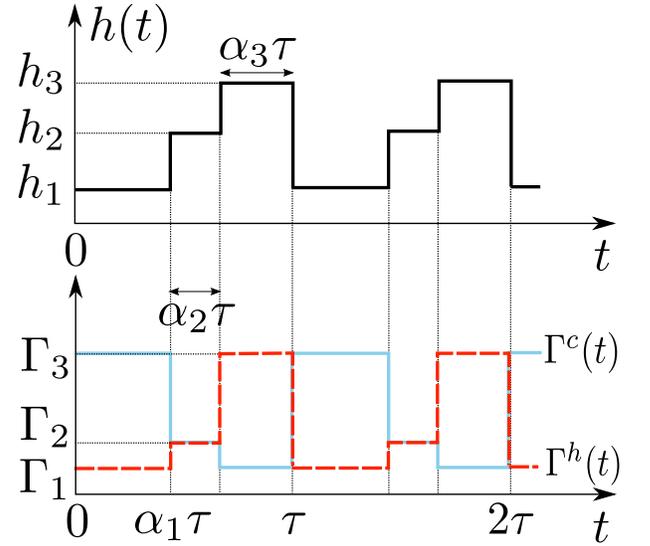}
\end{center}
\caption{(Top) External driving following a piecewise constant protocol of period $\tau $. The driving takes three different values $h_j$ with $j=1,2,3$ during three time intervals $\alpha_j \tau$, with $\alpha_1+\alpha_2+\alpha_3 = 1$. (Bottom) Time evolution of $\Gamma^\nu(t)$ indicating the coupling with the cold reservoir $\nu=c $ for the blue solid line and the hot reservoir $\nu = h$ for the red dashed line. Note that the reverse driving cycle is defined by $\hat h(t)=h(\tau-t)$ and $\hat \Gamma^\nu(t)= \Gamma^\nu(\tau-t)$.
 }
\label{fig6}
\end{figure}

\begin{figure}
\begin{center}
\includegraphics[width=\columnwidth]{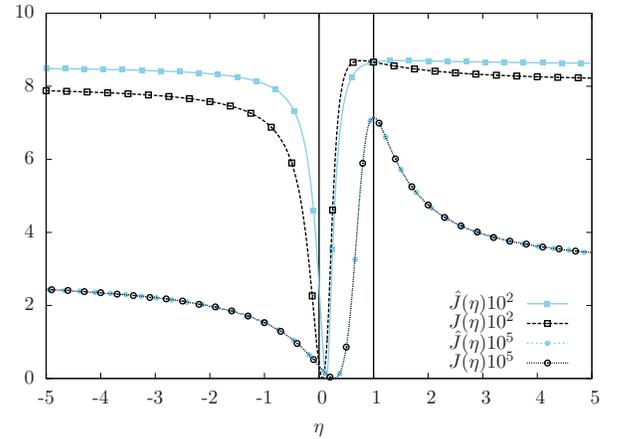}
\end{center}
\caption{efficiency LDF for a refrigerator (average behaviour) working far from equilibrium (squares) or close to equilibrium (circles) with the asymmetric driving cycle under time reversal of Fig.~\ref{fig6}. The empty (full) symbols are for the direct (reverse) driving. The kinetic is described by Fermi rates and the common parameters to all curves are $\alpha_1=0.6$, $\alpha_2 =\alpha_3=0.2$, $\tau=1$, $\Gamma_1=0.1$, $\Gamma_2=1$, $\Gamma_3=10$ and $\beta_h=1$. 
For far-from-equilibrium case, $h_1=0.5$, $h_2=1.5$, $h_3=2$ and $\beta_c = 1.5$. For close-to-equilibrium case, $h_1=1.45$, $h_2=1.5$, $h_3=1.55$ and $\beta_c = 1.05$. \label{fig11}}


\end{figure}

In order to illustrate the general results of section \ref{sec:AsymDriving}, we now consider the driving cycle depicted in Fig.~\ref{fig6} which is not symmetric under time-reversal. We see in Fig.~\ref{fig11} that as anticipated, the reversible efficiency is not the least probable anymore but lies at the intersection of the forward and time-reversed driving curves. This is only clearly seen far from equilibrium since the effect of the time-asymmetry of the driving vanishes as one approaches equilibrium as noted in the end of section \ref{sec:CloseEq}.

\section{Conclusion}\label{Conc}

We first summarize our results. 
Using the fluctuation theorem and assuming convexity of the currents LDF, we described the general properties of the LDF of efficiency fluctuations. 
Our conclusions hold for thermal and isothermal machines working arbitrarily far from equilibrium. 
We proved that the macroscopic efficiency defined as the ratio of average output power over average input power is the most probable efficiency. 
For general driving cycles the reversible efficiency is special in that the efficiency LDF of a machine subjected to a forward driving cycle and that of the same machine driven by the time-reversal protocol, coincide at that point. 
For machines operating at steady-state or subjected to time-symmetric driving cycles, the reversible efficiency is also the least likely efficiency as shown in Ref.\cite{Verley2014_vola}.
Close-to-equilibrium limit, the efficiency LDF is fully characterized by the response coefficients of the machine. Furthermore in this regime, machines subjected to a driving cycle or its time-reversed version display the same efficiency LDF.
We explicitly verified and illustrated our results by considering a two-level system machine subjected to piecewise constant driving protocols.
We finally also proposed a very efficient method to calculate the efficiency LDF directly from the cumulant generating function for the input and output currents. 

Nowadays stochastic quantities such as heat and work, have been measured in various small systems 
(e.g. biomolecules, systems of colloidal particles, polymers, quantum dots, single electron box). Hence their ratio, the stochastic efficiency, should be easily accessible experimentally. The statistical properties of the efficiency provide a much more accurate characterization of the performance of small machines then the macroscopic efficiency. In view of the high interest in the recent years for the study of finite-time thermodynamics at small scales,
we expect that the study of efficiency fluctuations will become a new paradigm in this field.    
Finally let us emphasize that the predictions of our theory for efficiency fluctuations provide a new way to verify the implications of the fluctuation theorem, which can be seen as the generalization of the second law for small systems.

\section*{Acknowledgement} 
This work was supported by the National Research Fund, Luxembourg under Project No. FNR/A11/02 and INTER/FWO/13/09. It also benefited from the COST action MP1209.
\appendix

\section{Work and heat CGF}\label{sec:CGFWQ}

We derive here the work and heat CGF for the two-state model of section \ref{sec:2statemodel} with the driving of Fig~\ref{fig0}. In this model, transitions between systems states require an instantaneous energy input or output, which corresponds to the heat exchanged with one of the reservoirs $\nu$. We use the label $\nu(t)$ to specify which reservoir caused the transition at time $t$. Energy conservation implies that the system energy change,
\beq
\Delta E(t) = W(t) + Q_c(t) + Q_h(t),
\eeq
can be expressed as the sum of the work provided by the driving
\beq
W(t) = - \int_0^t \D t' \dot h(t') \s(t'),
\eeq
and the heat provided by the reservoirs 
\beq
Q_\nu(t) = - \int_0^t \D t' h(t') \dot \s(t') \delta_{\nu,\nu(t)},
\eeq
where $\delta$ is the Kronecker symbol. 

For simplicity, we focus on the efficiency fluctuations of a refrigerator studying the statistics of work $W(t)$ and heat $Q_c(t)$. This implies no loss of generality: upon relabelling, we can also get the heat engine or heat pump efficiency fluctuations. The moment generating functions for work and heat at time $t$ conditioned on the final state $\s$ is defined by
\beq
G_{\s}(\gamma_1,\gamma_2,t)= \l e^{\gamma_1 W(t)+ \gamma_2 Q_c(t)} \delta_{\s,\s(t)} \r. \label{CondMGF}
\eeq
The one without conditioning is given by $G(\gamma_1,\gamma_2,t)=\sum_\s G_{\s}(\gamma_1,\gamma_2,t)$. 
The evolution of (\ref{CondMGF}) is ruled by the master equation
\beq
\partial_t G_{\s}(\gamma_1,\gamma_2,t) = \sum_{\s'=\pm 1} L^{(\gamma_1,\gamma_2)}_{\s,\s'}(h(t)) G_{\s'}(\gamma_1,\gamma_2,t), \label{eq:GenFuncEvo}
\eeq
where $\bm{L}^{(\gamma_1,\gamma_2)}$ is a $2$ by $2$ matrix dependent of $h$ with elements 
\bea
L^{(\gamma_1,\gamma_2)}_{\s,\s'}(h) &=& - \sum_{\nu=h,c} \s \s' \oh_{\nu}(h) e^{ -\beta^{\nu} \s' h + \gamma_2 h (\s'-\s) \delta_{\nu,l} } \nonumber \\ &&- \dot h \gamma_1 \s \delta_{\s,\s'}. \label{eq:generator}
\eea
This so called ``dressed'' generator of the evolution is equal to the master equation generator for the probability of the system states when $\gamma_1$ and $\gamma_2$ vanish. The asymptotic CGF of work and heat is related to the highest eigenvalue $\rho(\gamma_1,\gamma_2)$ of the propagator over one period of Eq.~(\ref{eq:GenFuncEvo}) written
\beq
\bm{Q} = \overrightarrow{\exp} \int_0^\tau \bm{L}^{(\gamma_1,\gamma_2)}(h(t)) \D t,
\eeq
where $\overrightarrow{\exp}$ stands for the time-ordered exponential. To see this, we write $g_{\s}(\gamma_1,\gamma_2)$ the right eigenvector of $\bm{Q}$ 
associated to $\rho(\gamma_1,\gamma_2)$ and $g(\gamma_1,\gamma_2) = \sum_\s g_{\s}(\gamma_1,\gamma_2)$ the sum of its components. Then, we have after $n$ periods
\bea
G(\gamma_1,\gamma_2,n \tau) &=& \sum_{\s,\s'} \left ( \bm{Q}^n\right )_{\s,\s'} g_{\s'}(\gamma_1,\gamma_2) \\ &=&  \rho(\gamma_1,\gamma_2)^n g(\gamma_1,\gamma_2),
\eea
leading to the asymptotic CGF of work and heat (per period) coming from the cold reservoir
\beq
\varphi_\mathrm{c}(\gamma_1,\gamma_2) = \lim_{n \ra \infty} \frac{1}{n} \ln G(\gamma_1,\gamma_2,n \tau) = \ln \rho(\gamma_1,\gamma_2). \label{eq:CGFJoint}
\eeq
In other word, we have to compute the matrix $\bm{Q}$ and look for its highest eigenvalue. This propagator follows from the product of four propagators for Eq.~(\ref{eq:GenFuncEvo}): the propagator between time $0$ and $\alpha^-\tau$ with the driving being $h^-$, the propagator over a unique time step during which occurs the transition from $h^-$ to $h^+$--  only the second line of the generator in Eq.~(\ref{eq:generator}) matter for this propagation-- the propagator between time $\alpha^-\tau$ and  $\tau$ with the driving being $h^+$ and, finally, the propagation over the time step of the transition from $h^+$ to $h^-$. These calculations have been described in more detailed in reference \cite{Verley2013_vol88} in the case of a modulated two-level system in contact with a unique heat reservoir but the calculations here are essentially the same. The final result for the CGF is 
\bea
\varphi_\mathrm{c}(\gamma_1,\gamma_2) = \ln \frac{  \mathrm{tr} \, \bm{Q} + \sqrt{ \left [ \mathrm{tr} \, \bm{Q}\right ]^{2} - 4 \det \bm{Q} }}{ 2} \label{eq:ScaledGenCumFunc},
\eea
which is a function of the determinant $\det \bm{Q} = z^+z^-$ with
$ z^\pm = \exp(-\tau \alpha^\pm k^\pm)$ and $k^\pm = 2 \sum_{\s,\nu} k_\nu(h^\pm,\s)$,
and of the trace
\begin{widetext}
\bea
\mathrm{tr} \, \bm{Q} &=& \sqrt{\frac{z^+z^-}{Z^+Z^-}} \left[ 1+ Z^+Z^- +(1-Z^+)(1-Z^-)  \frac{2\mathcal{C}-K^+K^-}{2K^+K^-} \right ], \quad \mbox{with}   \label{eq:trace}\\ 
\mathcal{C} &=& \underset{ \epsilon =\pm }{\sum_{ \mu,\nu = h,c}} \epsilon \, \omega_{\mu}^-\omega_{\nu}^+  \cosh ( \beta^{\mu \nu}_\epsilon h_0 -\beta^{\mu \nu}_{-\epsilon} a ) +\sum_{\mu,\nu =h,c}2 \omega_{\mu}^- \omega_{\nu}^+ \cosh [(\beta_+^{\mu\nu} -2\gamma_2 (\delta_{c,\mu}+\delta_{c,\nu}) +4\gamma_1)a-(\beta_-^{\mu\nu} -2\gamma_2 \varepsilon_{\mu \nu})h_0]. \nonumber
\eea
\end{widetext}
In these expressions, we have defined $\beta^{\mu\nu}_\pm =\beta_\mu\pm\beta_\nu$ for the sum and difference of temperatures, used the short notation $\oh_\nu(h^\pm)=\oh_\nu^\pm$ and introduced the Levi-Civita tensor for the heat reservoirs $\varepsilon_{cc} =\varepsilon_{hh} = 0$, $\varepsilon_{ch} = 1 $ and $ \varepsilon_{hc} =- 1$. We have also defined $ Z^\pm = \exp(-\tau \alpha^\pm K^\pm)$ and
\bea
K^\pm &=& \bigg \{ 4\sum_{\mu,\nu}  \omega_{\mu}^{\pm}\omega_{\nu}^{\pm} \big [ \cosh  h^\pm(\beta_-^{\mu\nu} -2\gamma_2 \varepsilon_{\mu \nu}) \nonumber \\
&&- \cosh  h^\pm\beta_-^{\mu\nu}  \big ] + (k^\pm)^2 \bigg \}^{1/2}. \label{eq:bigKdef}
\eea
We observe on equations.~(\ref{eq:ScaledGenCumFunc}-\ref{eq:bigKdef}) that $\lim_{\gamma_1 \ra \pm \infty } \varphi_\mathrm{c}(\gamma_1,0) = \pm 4a$ corresponds to the maximum slope of the CGF which is consistent with the fact that the extremal works value are $\pm 4a$. The heat exchanges are in principle unbounded, this corresponds, in the large $|\gamma_2|$ limit, to the fact that the CGF increases exponentially (no bounds on the slopes). As announced, the CGF of work and heat coming from the hot reservoir is define by
\beq
\varphi_\mathrm{h} (\gamma_1,\gamma_2) = \lim_{n \ra \infty} \frac{1}{n} \ln \l  e^{\gamma_1 W(t)+ \gamma_2 Q_\mathrm{h}(t)}  \r.
\eeq
and is obtained exchanging the labels c and h in equations.~(\ref{eq:ScaledGenCumFunc}-\ref{eq:bigKdef}). The CGF $\varphi_\mathrm{c}(\gamma_1,\gamma_2)$ and $\varphi_\mathrm{h}(\gamma_1,\gamma_2)$ provide all the required information to study the efficiency fluctuations of the three types of thermal machines as shown in Table \ref{tab:ThermalMachines}.

Evaluating the derivative of the CGF of work and heat derived at the origin, one obtain the average work per period:
\beq
\l w \r =\frac{ 8a \mathcal{Z} }{k^-k^+} \sum_{\mu \nu} \omega_{\mu}^{-} \omega_{\nu}^{+} \sinh (\beta_+^{\mu \nu } a - \beta_-^{\mu \nu} h_0) , \label{eq:avgWork}
\eeq
and the average heat per period coming from the cold reservoir:
\bea
\l q_\mathrm{c} \r &=& 4 \sum_{\epsilon=\pm} \left [ \frac{\mathcal{Z} }{k^\epsilon k^\epsilon} - \frac{\tau \alpha^\epsilon}{k^\epsilon} \right ] \omega_{c}^{\epsilon} \omega_{h}^{\epsilon} h^\epsilon \sinh{\beta_-^{ch} h^\epsilon} \nonumber \\ 
&-& \frac{4\mathcal{Z} }{k^-k^+} \bigg [ \sum_{\epsilon=\pm} \epsilon \omega_{c}^{\epsilon}\omega_{h}^{-\epsilon} h^\epsilon  \sinh (\beta_+^{ch}a+\epsilon \beta_-^{ch} h_0) \nonumber \\
&+& 2a \omega_{c}^{-} \omega_{c}^{+} \sinh 2 \beta_c a   \bigg], \label{eq:avgHeat}
\eea
where we have defined
\bea
\mathcal{Z} &=& (1-z^-)(1-z^+)/(1-z^-z^+).
\eea
As for the generating function, the average heat from the hot reservoir $\l q_\mathrm{h} \r$ is obtained by interchanging all the labels c with h.


\end{document}